\def\simgt{\lower.5ex\hbox{$\; \buildrel > \over \sim \;$}}
\def\simlt{\lower.5ex\hbox{$\; \buildrel < \over \sim \;$}}

\documentclass[useAMS,usenatbib]{mn2e}

\usepackage{graphicx}

\title[Dynamics of edge-on stellar discs]
{Structure and kinematics of edge-on galaxy discs --
V. The dynamics of the stellar discs.}
\author[M. Kregel, P.~C. van der Kruit \& K.~C. Freeman]
  {M. Kregel$^1$,
  P.~C. van~der~Kruit$^1$\thanks{E-mail: vdkruit@astro.rug.nl}
  and K.~C. Freeman$^2$ \\
  $^1$Kapteyn Astronomical Institute, University of Groningen,
  P.O.Box 800, 9700AV Groningen, the Netherlands\\
  $^2$Research School for Astronomy \&\ Astrophysics, Mount
  Stromlo Observatory, The Australian   National University, \\
  Private Bag, Weston Creek, 2611 Canberra, Australia}

\begin{document}

\date{Accepted. Received.}

\pagerange{\pageref{firstpage}--\pageref{lastpage}} \pubyear{2004}

\label{firstpage}

\maketitle

\begin{abstract}
In earlier papers in this series we determined
the intrinsic stellar disc kinematics of fifteen intermediate to late type
  edge-on spiral galaxies using a dynamical modeling
  technique. The sample covers a substantial range in
  maximum rotation velocity and deprojected face-on surface
  brightness, and contains seven spirals with either a boxy- or
  peanut-shaped bulge.
Here we discuss the structural, kinematical and dynamical properties.

From the photometry we find 
that intrinsically more flattened discs tend to have a lower
  face-on central surface brightness and a larger dynamical
  mass-to-light ratio. This observation suggests that at a constant
  maximum rotational velocity lower surface brightness discs have
  smaller vertical stellar velocity dispersions.

Although the individual uncertainties are large,
we find from the dynamical modeling 
that at least twelve discs are submaximal. The average disc
  contributes 53$\pm$4 percent to the observed rotation at 2.2 disc
  scalelengths ($h_{\rm R}$), with a 1$\sigma$ scatter of 15 percent. 
This percentage becomes somewhat lower when effects of finite disc 
flattening and gravity by the dark halo and the gas are taken into account.
Since boxy
  and peanut-shaped bulges are probably associated with bars, the
result suggests that at 2.2$h_{\rm R}$
the submaximal nature of discs is independent of barredness. The
possibility remains that very high surface brightness discs
are maximal, as these discs are underrepresented in our sample.
We confirm that the radial stellar
  disc velocity dispersion is related to the galaxy maximum rotational
  velocity. The scatter in this $\sigma-v_{\rm max}$ relation appears
  to correlate with the disc flattening, face-on central surface
  brightness and dynamical mass-to-light ratio. Low surface brightness
  discs tend to be more flattened and have smaller stellar velocity
  dispersions. The findings are consistent with the observed
  correlation between disc flattening and dynamical mass-to-light
  ratio and can generally be reproduced by
the simple collapse theory for disc galaxy formation.
  Finally, the disc mass Tully-Fisher relation is 
  offset from the maximum-disc
  scaled stellar mass Tully-Fisher relation of the Ursa Major
  cluster. This offset, $-$0.3 dex in mass, is naturally explained if
  the discs of the Ursa Major cluster spirals are submaximal.

surface brightness discs.

This paper has been accepted for publication by MNRAS
and is available in pdf-format
at the following URL:\\

http://www.astro.rug.nl/$\sim $vdkruit/jea3/homepage/paperV.pdf.

\end{abstract}


\end{document}